\documentclass[11pt]{article}
\usepackage{moriond,epsfig}

\def\APJ{{\em ApJ.} }
\def\MNRAS{{\em MNRAS} }

\def\be{\begin{equation}}
\def\ee{\end{equation}}
\def\bea{\begin{eqnarray}}
\def\eea{\end{eqnarray}}

\def\Msun{$M_\odot \;$}
\def\et{{\it et~al. }}

\def\spose#1{\hbox to 0pt{#1\hss}}
\def\simlt{\mathrel{\spose{\lower 3pt\hbox{$\mathchar"218$}}
     \raise 2.0pt\hbox{$\mathchar"13C$}}}
\def\simgt{\mathrel{\spose{\lower 3pt\hbox{$\mathchar"218$}}
     \raise 2.0pt\hbox{$\mathchar"13E$}}}
\def\simpropto{\mathrel{\spose{\lower 3pt\hbox{$\mathchar"218$}}
     \raise 2.0pt\hbox{$\propto$}}}

\begin{document}
\vspace*{4cm}
\title{A MICROLENSING SEARCH FOR COLD MOLECULAR CLOUDS IN VIRGO}

\author{ H. TADROS}

\secondaddress{Department of Physics, Astrophysics, Keble Road,\\
Oxford OX1 3RH, UK}

\author{ S. J. WARREN}

\secondaddress{Blackett Laboratory, Imperial College of Science, Technology
and Medicine, Prince Consort Rd, London, SW7 2BW, UK}

\author{ P. C. HEWETT}

\secondaddress{Institute of Astronomy, Madingley Road, Cambridge, CB3 OHA, UK}

\maketitle\abstracts{We report preliminary results from a first
season of photometric monitoring of 600 quasars behind the Virgo
galaxy cluster with the aim of detecting microlensing by the cluster
dark matter.  Our project is sensitive to dark objects of surface mass
densities down to $\sim 20 \,{\rm g \; cm}^{-2}$. We are thus capable
of detecting diffuse objects, such as cold molecular clouds, unlike
all Galactic microlensing surveys whose surface mass density limits
are $> 10^{4} \; {\rm g \; cm}^{-2}$.  The average optical depth to
microlensing of quasars through the central 30 sq. deg. of Virgo is
$\sim 1 \times 10^{-3}$. We report a null detection which implies that
less than half the dark matter in Virgo is in objects of mass
$\sim 10^{-5}$ solar masses, of surface mass density $> 20 \,{\rm g \;
cm}^{-2}$, at $90\%$ confidence.}

\section{Rationale}
The number of microlensing surveys underway is growing rapidly. Most
of these surveys aim to detect compact objects in the dark halo of our
own Galaxy or in the nearby M31 halo by monitoring stars in the LMC,
SMC, or M31 itself. These projects have proved very fruitful detecting
and placing limits on the dark matter in the Galaxy halo. The latest
results indicate that less than $20\%$ of the Galaxy halo is in the
form of MACHOs (see refs 1 and 2). There are, however, a number of
shortcomings of the current surveys. Firstly, one cannot measure the
distance to the MACHO and there is currently some controversy over
whether the events seen towards the LMC are due to Galactic dark
matter at all.  Secondly, the objects have to be extremely compact to
produce a microlensing effect at the distance of the LMC. Hence,
current surveys provide no constraint on the hypothesis that the dark
matter is composed of diffuse cold molecular hydrogen clouds (see
e.g. refs 3 and 4). Thirdly, source blending greatly complicates the
calculation of survey efficiencies.

We have begun a project which addresses these issues and provides
completely new information on the nature of dark matter in galaxy
clusters. We are monitoring 645 quasar candidates behind the Virgo
galaxy cluster in order to detect dark objects in the cluster in
the mass range $5 \times 10^{-7}$ \Msun -- $5 \times 10^{-3} $
\Msun. The upper mass limit is set by the length of the monitoring
program and the lower limit by the frequency of the monitoring.
Quasar source size can also have an effect on the lower mass limit. If
the quasar's angular size is larger than the angular diameter of the
Einstein ring of the lens, the magnification drops significantly and
the event would not be detected. If all the mass in the Virgo cluster
were in dark objects within the above mass range then we can expect to
see significant numbers of microlensing events during our monitoring
program (see ref 5 for details).

Several features of our monitoring program deserve to be
emphasised. Firstly, if we see a microlensing event we have a
reasonably accurate measure of the distance to the MACHO, allowing us
to calculate its mass.  Secondly, we can detect objects down to
surface mass densities of a few tens of $ {\rm g \; cm}^{-2}$ as these
would act as point mass lenses when placed in Virgo. Such objects, for
example cold molecular gas clouds, are accessible to no other current
microlensing survey.  Thirdly, because we are monitoring $\sim 600$
objects over a 30 sq. deg. field, we do not have blending of source
images and the associated problems that blending causes.

One potential worry for a survey such as ours is the fact that quasars
are intrinsically variable. If quasar variability is large on the time
scales over which we are monitoring we would have to set our
microlensing magnification threshold so high that the expected number
of events would become very low. To date, little systematic quasar
monitoring has been carried out on time scales matching those of our
project, thus we have been careful to examine the feasibility of our
project. As reported in this article, our pilot project proves that
intrinsic quasar variability is not a serious problem on these time
scales.  A by-product of this microlensing survey is the information
provided on the variability of quasars over these timescales, which
will lead to constraints on models of quasar fuelling mechanisms.

\section{Observations and data analysis}
During the period 1999 February - June we obtained 28 R-band Schmidt
plates of the central 30 sq. deg. of the Virgo cluster. The data are
complete to an R-band magnitude of $20$. Using previous plates of the
same area in the U and B bands we identified $645$ candidate quasars
according to their colour. Of these we expect $\simlt 10 \%$ to be
contaminant stars. Figure 1 shows $2$ typical quasar light curves
extracted from this data set.

It can be seen qualitatively that the quasar candidates vary little
during the monitoring period. We have quantified this by comparing the
level of variability in the quasar candidates with the variability of
$3000$ stars in our field. The candidate quasars are statistically no
more variable than the stars. This confirms the feasibility of using
quasars for microlensing surveys for monitoring periods of a few
months.  Our simulations show that with this low level of variability
we can set a low detection threshold, and identify microlensing events
down to a peak magnification of 0.46 mag.

\begin{figure}
\centering
\begin{picture}(90,90)
\includegraphics{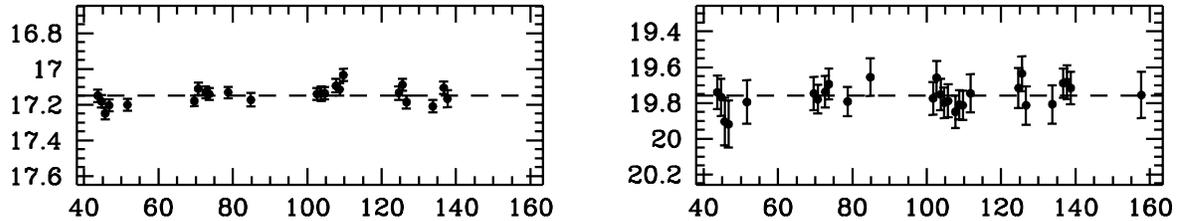}
\end{picture}
\caption{Typical light curves for quasar candidates monitored over a
period of 113 days, plotting R mag. against 1999 day.}
\end{figure}

We have run a matched filter analysis on each of the $645$
candidate-quasar light curves. The filter used was a MACHO light curve
of peak amplification $0.6$ mag. We used a range of values for the
event time-scale and the time of peak magnification. For each quasar
light curve the event with the highest $\Delta \chi^{2}$ for a match
to a MACHO event was retained.  To this best event for each quasar, we
performed a four parameter MACHO light curve fit.

To define selection criteria to pick out likely lensing events, we ran
the same set of procedures on the light curves of the 3000 control
stars in the field -- where we do not expect any real microlensing
events.  Figure 2 shows the distribution of points in the plane of
$\Delta \chi^{2}$ for the four parameter MACHO fit against the fitted
impact parameter $b_{min}$. Quasar candidates are shown by crosses and stars as
stars. We define the following criteria for an event to be considered
as a serious microlensing candidate:\\ $\bullet$ $\Delta \chi^{2} >
100$\\ $\bullet$ ${\mbox{Impact parameter}} \; b_{min}< 0.8$\\ These criteria
leave two events in the candidate region to be checked. Neither of
these proved to be real events: one is a known highly variable BL Lac
object and the other a probable stellar RR Lyrae contaminant.  From
this null detection we can now put constraints on the nature of dark
matter in Virgo.


\begin{figure}
\centering
\begin{picture}(215,215)
\includegraphics{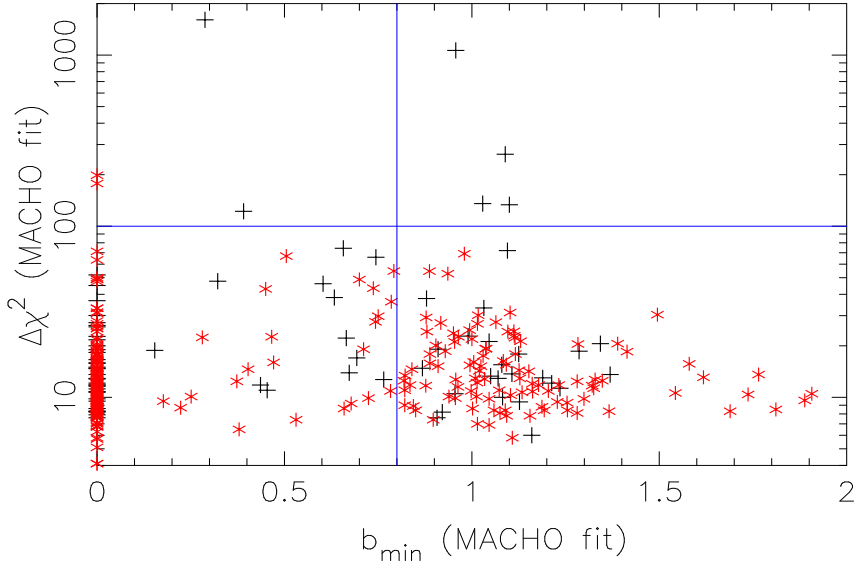}
\end{picture}
\caption{Plot illustrating microlensing candidate
selection. For the best fitting MACHO event in each quasar (crosses)
and star (asterisks) light curve we plot the $\Delta \chi^{2}$ against the
fitted impact parameter. Our candidate selection criteria, $\Delta
\chi^{2} > 100$, $b_{min} < 0.8$, are designed to exclude all events in
the stellar light curves (where we expect no real microlensing
events).}
\end{figure}

\section{Detection Efficiency and Predicted Number of Events}
To calculate the detection efficiency of the survey
we have generated artificial microlensing events and added them
to the 3000 stellar light curves. For each event time-scale we have
generated $10,000$ artificial light curves with impact parameters
distributed uniformly between $0$ and $1$ Einstein ring radius and
added them to the stellar light curves, placed at random dates.  We
then applied our matched filter analysis, the fitting routine, and
the selection criteria to these artificial light curves and measured how
many of the artificial events were recovered. Our detection efficiency
is shown as a function of event time-scale on the left hand side of
Figure 3.

Using the detection efficiency we can calculate the number of
microlensing events we would expect to see in our experiment if the
dark matter in Virgo were in the form of dark objects of a particular
mass. We assume that the Virgo cluster can be modelled as an
isothermal sphere with a one-dimensional velocity dispersion of
$\mathrm {673\, km\, s^{-1}}$.  Our expected number of events is shown
on the right hand side of Figure 3 and peaks at $5$, if all the dark
mass in Virgo were in the form of $10^{-5}$\Msun objects.

\begin{figure}
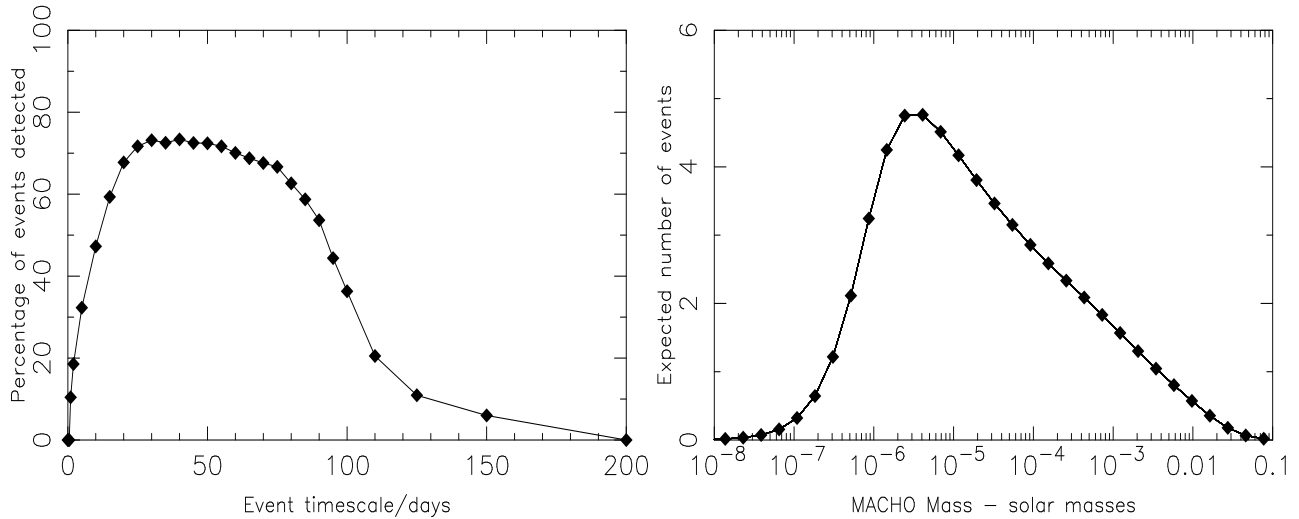

\centering
\begin{picture}(190,190)
\includegraphics{that.ps}
\includegraphics{npred.ps}
\end{picture}
\caption{The left hand side of the figure shows the detection
efficiency of our microlensing experiment as a function of MACHO event
time-scale. The right hand side shows the expected number of MACHO
events as a function of mass, assuming that all of Virgo is composed
of objects of that mass.}
\end{figure}

\section{Conclusions and future work}
From our null detection, and the fact that we would have expected $5$
events for lenses of mass $1 \times 10^{-5}$ \Msun, we can use Poisson
statistics to put constraints on the mass in Virgo in the form of
dark objects of this mass. The result of this pilot project is that less
than $1/2$ the mass in Virgo can be in the form of $10^{-5}$ \Msun
objects at $90\%$ confidence.

The most important thing we have learnt from the pilot project is that
a quasar monitoring microlensing project is technically feasible and
we hope, over the next 2 years, to acquire of order $100$ more plates
of Virgo. Our simulations indicate that this will allow us to improve
our constraints on the nature of dark matter in Virgo by a factor
$5-10$. A second season of monitoring Virgo started in February 2000.

We would also like to extend the experiment by monitoring other
clusters - especially the Perseus cluster in the North as this is 
the most massive nearby cluster. Alternatively, another strategy would be to
monitor several more distant clusters (thereby avoiding the
need for a very wide field of view).  In the long term we would also
like to monitor a control field of quasars with no
intervening massive cluster in order to precisely quantify quasar
variability.

\section*{Acknowledgements}
We are immensely grateful to the Schmidt observing team: Fred Watson,
Malcolm Hartley, Russell Cannon, Paul Cass, Ken Russell and Delphine
Roussiel for taking all the photographic plates of Virgo used in this
analysis.

\end{document}